\documentclass[useAMS,usenatbib]{mn2e}

\title[Space-charge-limited flow]{The effect of photo-electric absorption on 
space-charge limited flow in pulsars}
\author[P. B. Jones]{P. B. Jones\thanks{E-mail:p.jones1@physics.ox.ac.uk}\\
Department of Physics, University of Oxford, Denys Wilkinson Building,\\
Keble Road, Oxford OX1 3RH, England}

\begin{document}

\date{}

\pagerange{\pageref{firstpage}--\pageref{lastpage}} \pubyear{}

\maketitle

\label{firstpage}

\begin{abstract}
Photo-electric absorption of blackbody photons is an important process which limits the 
acceleration of ions under the space-charge limited flow boundary condition at the 
polar caps of pulsars with positive corotational charge density.  Photo-electric 
cross-sections in high magnetic fields have been found for the geometrical conditions 
of the problem, and ion transition rates calculated as functions of the surface 
temperatures on both the polar cap and the general neutron-star surface.  The general 
surface temperature is the more important and, unless it is below $10^{5}$ K, limits 
the acceleration electric field in the open magnetosphere to values far below those 
needed either for electron-positron pair creation or slot-gap X-ray sources.  But such 
ion beams are unstable against growth of a quasi-longitudinal Langmuir mode at rates 
that can be observationally significant as a source of coherent radio emission.
\end{abstract}


\section{Introduction}

Two previous papers have given the results of a study of reverse-electron flow in 
isolated neutron stars with positive polar-cap corotational charge density $\rho 
_{GJ}$, in which free emission of ions maintains the boundary condition for 
space-charge limited flow. The sign of this quantity depends on the relation between 
rotation spin ${\bf \Omega}$ and polar magnetic flux density ${\bf B}$. We refer to 
Goldreich \& Julian (1969) for its Euclidean-space definition, and to Muslimov \& 
Tsygan (1992) and Muslimov \& Harding (1997) for the general-relativistic modification 
used here.
Most publications have assumed the parallel case, ${\bf \Omega}\cdot{\bf B} > 0$, with 
$\rho _{GJ} < 0$ and electron acceleration.  But there does not appear to be any reason 
why the antiparallel case with $\rho _{GJ} > 0$ and ion and positron acceleration 
should not be present in the neutron-star population.  The two earlier papers (Jones 
2010a, 2011; hereafter Papers I and II) considered in some detail the consequences of 
proton formation in the electron-positron showers produced by the reverse-electron flux 
at the polar cap and its relation with observed radio-pulsar phenomena.  But an 
important omission was the effect of reverse-electron flow on the electrostatics of 
particle acceleration under the space-charge limited boundary condition, and this is 
the subject of the present paper.

In this paper, the subscripts $\parallel$ and $\perp$ are with reference to the local 
magnetic field direction.
Under the surface electric-field boundary condition ${\bf E}_{\parallel}\neq 0$
assumed in the classic paper of Ruderman \& Sutherland (1975), reverse-electron flow 
does not modify ion or positron acceleration in any important way because the surface 
value of the acceleration field ${\bf E}_{\parallel}$ is unconstrained. However, the 
presence of a reverse-electron negative charge density under the surface boundary 
condition ${\bf E}_{\parallel} = 0$ of space-charge limited flow can, in certain cases, 
reduce the acceleration field to values so low that there is no possibility of 
electron-positron pair creation within the open magnetosphere above the polar cap or of 
a slot-gap X-ray source.
Photo-electric absorption of polar-cap and neutron-star surface blackbody photons by 
accelerated ions is the source of the reverse-electron flux and the process is 
analogous with electron-positron pair creation in its effect on the acceleration field.
Heating of the polar-cap surface by a reverse flux of photo-electrons has been 
considered previously by Cheng \& Ruderman (1977), also by Muslimov \& Tsygan (1987) in 
relation to an instability of a liquid surface under the boundary condition ${\bf 
E}_{\parallel} \neq 0$.  The present paper is concerned not with heating effects but 
with the electrostatic consequences of the negative charge density that 
reverse-electron flow implies. 

There have been many calculations of the photo-electric cross-section for atomic 
hydrogen in the high magnetic fields and temperatures of model neutron-star 
atmospheres.  We refer to Potekhin \& Pavlov (1997) for a critical survey of this work 
and to Potekhin, Pavlov \& Ventura (1997) for a study of the adiabatic approximation 
used by many authors.  Although model atmospheres  containing low-$Z$ elements have 
been constructed (see, for example, Mori \& Ho 2007),
there appear to be no published explicit photo-electric cross-sections for such 
elements in high magnetic fields. Also, the intervals of electron separation and photon 
energies relevant to the accelerated ions in the present work differ from those of 
neutron-star atmospheres.  But the requirements of this paper have some simple features 
compared with the latter problem.  It is not necessary to consider the complexities of 
thermal ion motion transverse to the field and, in particular, the photon momentum 
component ${k_{\perp}}$ is such that in relation to the electron cyclotron radius 
$r_{B}$, the condition $k_{\perp}r_{B} \ll 1$ is always satisfied.  Hence simple 
expressions, valid only in this limit,  have been derived and are given in Appendix A. 
The zero-field cross-sections have been computed extensively (see, for example, the 
work of Reilman \& Manson 1979)  and in approximations much superior to those of 
Appendix A, but nevertheless we show that use of the high-magnetic field expressions is 
essential in this paper.

The region of open magnetosphere above the polar cap is assumed to have a circular 
cross-section of radius $u_{0}(z)$, where $z + R$ is the coordinate on the magnetic 
axis and $R$ is the neutron-star radius. The acceleration field has two components. The 
first, present at very low altitudes $z \ll u_{0}(0)$ above the polar-cap surface, is 
the inertially-generated field described by Michel (1974). The second, and much the 
more important, contribution is produced by the Lense-Thirring effect modification to 
the corotational charge density first recognized by Muslimov \& Tsygan (1992) and 
extends over moderately large distances $z \sim R$.  The computational results of this 
paper are contained in Table 1, which gives the charge evolution of typical ions as 
they are accelerated.

The photon source is modelled as a polar cap of radius $u_{0}(0)$ and proper-frame 
temperature $T_{pc}$ embedded in a spherical neutron-star surface of uniform 
temperature $T_{s}$.  We find that the latter temperature is much the more significant 
but, unfortunately, it is essentially unknown in the case of the general isolated 
neutron-star population.  Blackbody emission has been detected  reliably in only a 
small number of pulsars with ages less than $10^{6}$ yr (see the review of Yakovlev \& 
Pethick 2004) and even then there remains the problem of deciding whether the source is 
the polar cap or the whole surface.

Section 2 considers the simplest possible acceleration process: one that is 
time-independent and in a common state of plasma composition over the whole polar cap.  
It is shown that, in this case, acceleration is inadequate to produce secondary 
electron-positron cascades except for values of $T_{s}$ that are too small to be 
observable. The relation between this result and earlier work on space-charge limited 
flow in the ${\bf \Omega}\cdot{\bf B} < 0$ case
(Papers I and II) is discussed in section 3. The flux of ions that are relativistic but 
not ultra-relativistic is unstable against growth of a quasi-longitudinal Langmuir mode 
at rates that appear large enough to produce Langmuir solitons and strong  turbulence 
(Weatherall 1997, 1998; Melrose 2000; Melikidze Gil \& Pataraya 2000; Asseo \& Porzio 
2006).

\section{Acceleration and photo-electric absorption}

Solutions for the charge density $\rho$, for $\rho _{GJ}$ and the electrostatic 
potential $\Phi$ have been obtained  for space-charge limited flow in a dipole field by 
Muslimov \& Harding (1997) and by Harding \& Muslimov (2001).  These authors considered 
electrons, for which the inertial component of the acceleration field is negligible.  
For ion acceleration, the inertial component, though less important than the 
Lense-Thirring induced term should be included.  The expressions given by Muslimov \& 
Harding are  complicated and therefore we shall make use of the boundary condition and 
geometrical form of the open flux-line magnetosphere to obtain a very simple 
approximate expression for $\Phi$ which is adequate for the purposes of this paper.  We 
calculate the photo-electric transition rate for an ion accelerated to Lorentz factor 
$\gamma$ at altitude $z$ above the magnetic pole.

There is little information about electron separation energies for multiply ionized 
atoms in the magnetic fields, $10^{12} < B < 10^{13}$ G of interest here.  The most 
recent calculations (Medin \& Lai 2006) are restricted to atomic numbers $Z = 6$ and 
$26$ and further, in the latter case to $B \geq 10^{14}$ G.
We therefore adopt the expression used previously (Jones 1981) for the separation 
energy $\epsilon _{l}$,
\begin{eqnarray}
\epsilon _{l} = 0.35B^{1/2}_{12}\frac{Z-l}{\sqrt{2l + 1}}\hspace{5mm}{\rm keV},
\end{eqnarray}
in which $l = 0,1,2,..$ is the electron spatial degeneracy quantum number. When summed 
over all $l$, equation (1) gives agreement with the directly computed total energies of 
atoms with atomic numbers $Z = 10$ and $20$
(see Jones 1985) that is adequate for the work of this paper.

The inertial component of the acceleration field can be found at small $z \ll u_{0}(0)$ 
by means of a one-dimensional approximation to the Poisson equation which neglects the 
boundary condition on the cylindrical surface $u = u_{0}$ separating open and closed 
flux lines,
\begin{eqnarray}
E_{\parallel} \approx E_{i} = \left(\frac{8\pi \rho 
_{GJ}Mc^{2}}{\tilde{Z}e}\right)^{1/2},
\end{eqnarray}
for ions of mass $M$ and charge $\tilde{Z}$.  The field increases from the boundary 
condition ${\bf E}_{\parallel} = 0$ to this constant value in a very small distance
\begin{eqnarray}
z_{i} \approx \left(\frac{Mc^{2}}{18\pi\rho _{GJ}\tilde{Z}e}\right)^{1/2}.
\end{eqnarray}
Photo-electric absorption of blackbody photons modifies $E_{\parallel}$ by introducing 
the negative charge density of the electrons accelerated toward the polar-cap surface.  
To estimate their effect, we can assume that the transitions all occur at a unique 
altitude $z = h$, increase the ion charge from $\tilde{Z}$ to $Z_{h}$, and reduce the 
field to $E_{\parallel} = 0$ at $z = h$.  Then,
\begin{eqnarray}
h \approx \frac{3}{2}z_{i}\frac{\tilde{Z}}{Z_{h} - \tilde{Z}},
\end{eqnarray}
and the ion Lorentz factor at $z = h$ is,
\begin{eqnarray}
\gamma _{h} \approx 1 + \frac{1}{2}\frac{\tilde{Z}}{Z_{h} - \tilde{Z}}.
\end{eqnarray}
The electron chemical potential in the upper, low-density, regions of the atmosphere at 
the polar cap surface is,
\begin{eqnarray}
\mu _{e} = k_{B}T\ln\left(\pi r^{2}_{B}N_{e}\left(\frac{2\pi \hbar^{2}}
		{mk_{B}T}\right)^{1/2}\right),
\end{eqnarray}
at electron number density $N_{e}$.
We have estimated $\tilde{Z}$ by comparing the value of $\mu _{e}$ on the surface of 
last scattering in the atmosphere prior to acceleration with the
$\epsilon _{l}$ given by equation (1).  On this basis, and because the dependence of 
$\mu _{e}$ on $B$ is only logarithmic, we have adopted the condition $\epsilon _{l} = 
26k_{B}T$ to obtain the $\tilde{Z}$ used in Table 1.
The polar-cap radius is that given by Harding \& Muslimov (2001),
\begin{eqnarray}
u_{0}(0) = \left(\frac{2\pi R^{3}}{cPf(1))}\right)^{1/2},
\end{eqnarray}
in which $P$ is the rotation period and $f(1) = 1.368$ for a neutron star of mass 
$1.4M_{\odot}$ and radius $R = 1.2 \times 10^{6}$ cm.

The inertial component of $E_{\parallel}$ ceases to be significant at $h \sim 10^{3}$ 
cm.  In principle, calculation of the electric potential $\Phi$ at $z > h$ requires the 
solutions given by Harding and Muslimov.  But at $z \gg u_{0}(0)$ a very simple 
expression can be found on the basis that the open flux lines are contained within a 
narrow cylindrical tube whose cross-sectional area increases only slowly with $z$.  
Given the boundary condition $\Phi = 0$ on the surface $u = u_{0}$ and that the radial 
electric field is $E_{\perp} \gg E_{\parallel}$, application of Gauss's theorem to a 
section of the tube at altitude $z$ gives the approximation,
\begin{eqnarray}
\Phi = \pi\left(u^{2}_{0}(z) - u^{2}\right)\left(\rho(z) - \rho _{GJ}(z)\right),
\end{eqnarray}
in terms of the difference between the charge density $\rho(z)$ and the 
Goldreich-Julian charge density $\rho _{GJ}(z)$. We refer to Harding \& Muslimov (2001) 
for this quantity at polar coordinate $r = \eta R$ and adopt the expression,
\begin{eqnarray}
\rho - \rho _{GJ} = \frac{\kappa Bf(\eta)}{cP\alpha\eta^{3}f(1)}
	\left(1 - \frac{1}{\eta^{3}}\right)\cos\psi,
\end{eqnarray}
in which we have neglected a second term containing $\sin\psi$ which is small at 
altitudes $\eta$ of order unity.  Equation (9)
assumes a dipole field with surface value $B$.  Here, $\kappa$ is the dimensionless 
Lense-Thirring factor and $\psi$ is the angle between ${\bf \Omega}$ and ${\bf B}$.
This expression is a satisfactory approximation at $z > u_{0}(0)$ and $z < 3R$.  At 
these altitudes, the function $f(\eta)/\alpha f(1)$ including the red-shift factor 
$\alpha$ can be well-approximated by unity and is a slowly varying function of $\eta$.

Equations (8) and (9) define the acceleration of an ion and give the relation between 
the altitude $\eta$ and the Lorentz factor $\gamma$ and we restrict transition rate 
calculations to points on the magnetic axis ($u = 0$).
The neutron-star surface is assumed to be Lambertian.  Then unit element of surface 
area at a point which subtends an angle $\zeta$ with the magnetic polar axis produces a 
photon number flux,
\begin{eqnarray}
J_{0}(\omega _{0},\theta) = \frac{\omega^{2}_{0}n(\omega _{0})}{4\pi^{3}c^{2}R^{2}}
	\frac{\cos(\zeta + \theta)}{1 + \eta^{2} - 2\eta\cos\zeta},
\end{eqnarray}
at altitude $z = r - R$ on the axis, in which the photon occupation number is $n(\omega 
_{0})$ for photons of angular frequency $\omega _{0}$.  The photon momenta from this 
element are at an angle $\theta$ with the magnetic axis.  Lorentz transformation 
parallel with ${\bf B}$ to the rest frame of an ion on the axis moving outward with 
Lorentz factor $\gamma$ gives the flux in that frame,
\begin{eqnarray}
J(\omega,\chi) = \gamma J_{0}(\omega _{0},\theta)(1 - \beta\cos\theta),
\end{eqnarray}
where $\omega$ and $\chi$ replace $\omega _{0}$ amd $\theta$, and $\beta$ is the ion 
velocity in units of $c$.  The photo-electric transition rate transformed back to the 
neutron-star frame at an altitude $\eta - 1$ on the magnetic axis is,
\begin{eqnarray}
\Gamma _{l} = \frac{1}{\gamma}\int^{\zeta _{max}}_{0}2\pi R^{2}\sin\zeta d\zeta
\int^{\infty}_{0}d\omega J(\omega,\chi)\sigma(B,\omega,\chi),
\end{eqnarray}
in which the upper limit of integration is given by $\cos\zeta _{max} = 1/\eta$
and the integral is in two sectors; the polar cap with temperature $T_{pc}$ and radius 
given by equation (7), and the remaining visible neutron-star surface of temperature 
$T_{s}$.
The cross-section is $\sigma  = (\sigma^{a}_{0} + \sigma^{a}_{1} + \sigma^{b}_{1})/2$.  
Here, the subscripts denote the Landau quantum number of the final electron.  The 
photon polarization is denoted by the superscripts $a,b$, respectively perpendicular to 
or parallel with ${\bf k}\times {\bf B}$.
General-relativistic corrections to photon propagation, including the red-shift factor, 
have been neglected here in equations (10) - (12).  Equation (12) gives the transition 
rate as a function of the coordinate $\eta$ on the magnetic axis and the Lorentz factor 
$\gamma$ of the ion. Owing to the geometrical restriction of equation (12) to points on 
the magnetic axis, transition rates for ions accelerated off-axis at finite $u(z)$ 
would be in error to a modest extent for the polar-cap contribution, but for the much 
more significant whole-surface term, the error is negligible.
We refer to Appendix A for details of the individual cross-sections.  Approximations 
and sources of error are also discussed there, as is the comparison with zero-field 
cross-sections.

The integration in equation (12) is over the polar cap with temperature $T_{pc}$ and 
the remaining visible part of the neutron-star surface at temperature $T_{s}$. From the 
transition rates and the relations between $\gamma$ and $z$ given by equations (2) - 
(5), (8) and (9),
two optical depths have been determined for three values each of nuclear charge and 
polar-cap magnetic flux density.  These are,
\begin{eqnarray}
\frac{1}{c}\int^{h}_{0}\Gamma _{l}(z)dz
\end{eqnarray}
which defines $Z_{h}$ for a given $\tilde {Z}$. Its dependence on $T_{s}$ is negligibly 
small principally because the factor $\cos(\zeta + \theta) \approx 0$. The second 
optical depth is,
\begin{eqnarray}
\frac{1}{c}\int^{z_{max}}_{h}\Gamma _{l}(z)dz
\end{eqnarray}
which gives $Z_{\infty}$ and is almost completely independent of $T_{pc}$ owing to the 
unfavourable nature of the Lorentz transformation for polar-cap photons at $z \gg 
u_{0}(0)$. In both cases, integer charges are determined by whether or not the optical 
depth concerned is greater or less than unity.  This is usually obvious because optical 
depths decrease rapidly with increasing ion charge and electron separation energy.

Equations (4) and (5) are based on the effect of reverse-electron negative charge 
density at $z \ll u_{0}(0)$. Photo-electric absorption at $z \gg u_{0}(0)$ also 
introduces  a reverse-electron flux and negative charge density in any interval of $z$ 
in which there is a finite $E_{\parallel}$.  Thus $\rho u^{2}_{0}$ for 
ultra-relativistic ions is not constant but
is a function of $z$ determined by the local transition rate,
\begin{eqnarray}
\frac{1}{\rho u^{2}_{0}}\frac{\partial(\rho u^{2}_{0})}{\partial z} = 
\sum _{l}\frac{2\Gamma _{l}}{cZ(z)}.     \nonumber
\end{eqnarray}
and so reduces the acceleration field obtained from equation (8).  There can be 
acceleration in any interval of $z$ only if photo-electric rates are such that 
$\partial\Phi/\partial z < 0$ and this condition leads at once to the inequality,
\begin{eqnarray}
\sum _{l}\frac{2\Gamma _{l}}{cZ(z)} < \kappa\frac{\partial}{\partial z}
\left(1 - \frac{1}{\eta^{3}}\right),
\end{eqnarray}
in which the fractional error in the right-hand term is of order $\kappa$ relative to 
unity, and $Z(z)$ is the local value of the ion charge.
Because the principal aim of this calculation is to see if there can be sufficient 
acceleration to produce electron-positron pair creation, we adopt the following 
procedure.  We scale $\Phi$ so as to give an ion Lorentz factor $\gamma = 500$ at 
$z_{max}$, that is, a potential difference of about $10^{3}$ GeV.  Then the lowest ion 
charge that has less than unit optical depth in the interval $h < z < z_{max}$ is 
defined as $Z_{\infty}$.  But if this acceleration is to be possible, photo-electric 
rates must not exceed the limit defined above.  On integrating (15), we find that 
$Z_{\infty}$ must satisfy the  approximate inequality $Z_{\infty} - Z_{h} < \kappa 
Z_{\infty}/2$ for consistency.  This condition is maintained by an equilibrium such 
that the acceleration field satisfying Poisson's equation adjusts to those values that 
give the ion acceleration and photo-electric transition rates necessary for 
self-consistency.  To enlarge on this, we can note that if photo-ionization were 
complete at an altitude $z_{c}$, acceleration of the bare nucleus would continue at $z 
> z_{c}$, the potential difference remaining for this being approximately $\Phi(z) - 
\Phi(z_{c})$.  But integration of (15) for this case gives the approximate constraint,
\begin{eqnarray}
Z - Z_{h} < \frac{\kappa Z}{2}\left(1 - \frac{1}{\eta^{3}_{c}}\right), \nonumber
\end{eqnarray}
so that values of $Z - Z_{h}$ for which continued acceleration is possible are severely 
constrained.  The equilibrium we have mentioned above means that the acceleration field 
given by Poisson's equation adjusts at all $z$ to that required for photo-ionization 
transition rates no larger than those satisfying the inequality (15).
Reference to Table 1 shows that continued acceleration over a potential difference 
approaching the maximum given by equations (8) and (9) is not possible except for low 
$B$ and high values of $T_{pc}$.

The optical depth integration is continued to $z_{max} = 3R$ although the major 
contribution is from much lower altitudes. But the local magnetic flux density is much 
reduced near $z_{max}$ and therefore the separation energy used is the greater of that 
given by equation (1) or a zero-field expression given by a Rydberg formula in terms of 
the screened nuclear charge and the appropriate principal quantum number for the 
electron shell.

The Table shows that $Z_{h}$ differs little from $\tilde{Z}$ for any values of $B$ and 
$T_{pc}$.  There are two reasons for this.  Firstly, the cross-section for transitions 
to the lowest Landau state, the only final electron state accessible at low values of 
$\gamma$, contains the factor $\sin^{2}\chi$, which is always small.  Thus use of the 
zero-field cross-section in this region would be quite incorrect.  Secondly, as 
previously explained in relation to equations (4) and (5), the reverse-electron 
negative charge density itself limits acceleration at $z < h$.  Most photo-electric 
absorption occurs at $z \gg u_{0}(0)$,
is $T_{s}$-dependent, and consists of transitions to the $n = 1$ Landau state whose 
cross-sections contain $\cos^{2}\chi$. These, of course, require acceleration to higher 
Lorentz factors.

The estimate of the Lense-Thirring parameter used by Muslimov \& Harding (1997) is 
$\kappa = 0.15$. 
Values of $Z_{\infty}$ are given for neutron-star surface temperature $T_{s} = 1.0 
\times 10^{5}$ K in column 6 and by reference to the integrated form of (15)
show that $Z_{\infty} - Z_{h}$ is such that acceleration is not always possible even at 
this low temperature.  At $2.0\times 10^{5}$ K, acceleration occurs only in the limited 
number of cases in which high values of $T_{pc}$ give $Z_{h} = Z$. Because $Z_{h}$ 
differs little from $\tilde{Z}$, we can see that the basic uncertainty is in the latter 
quantity.  Estimates more refined than our condition derived directly from equations 
(1) and (6) might give lower or higher values of $\tilde{Z}$.  In the former case, 
values of $Z_{h}$ would also be smaller, hence the possibility of significant 
acceleration would be reduced.  In the latter, $Z_{\infty} - Z_{h}$ would be reduced. 
However, if photo-electric ionization is complete at $z = h$, the energy flux of 
reverse electrons is very limited.  Typically, for $Z_{h} - \tilde{Z} = 1$, equation 
(5) shows that the reverse electron energy input to the polar cap is no more than $\sim 
20$ GeV per ion, giving only low rates of pair creation by inverse Compton scattering 
or by (n,$\gamma$) reactions.

The values of polar-cap magnetic flux density considered in Table 1 were chosen so as 
include the median and mean values of $\log _{10}B$ for the Wang, Manchester \& 
Johnston (2007) catalogue of pulsars that exhibit nulls. For fields much smaller than 
$10^{12}$ G, the approximation $k_{\perp}r_{B} \ll 1$
made in Appendix A fails, Landau level spacings become small, and the photo-electric 
cross-section approaches the zero-field value.  Thus transition rates remain high. On 
the basis of Table 1, we find, with some confidence, that the Muslimov-Tsygan 
acceleration field is in general so reduced by electron production that 
electron-positron pair creation is not possible. In this state, which we assume to be 
time-independent, there is an equilibrium between the photo-electric transition rates 
and the ion acceleration needed to produce them. In many cases, ion Lorentz factors at 
$z_{max}$ may differ little from $\gamma _{h}$.  Our finding is limited, of course, to 
neutron stars whose whole-surface temperature is $T_{s} \geq 10^{5}$ K. But a 
temperature of $10^{5}$ K is problematical from all points of view.  It is unobservable 
in a compact source given present techniques and the energy flux needed to maintain it 
is small and easily supplied by many possible dissipative processes. Also, measured 
photon spectra do not always allow the unambiguous separation of polar cap and 
whole-surface radiation.

\begin{table}
\caption{For a range of values of polar-cap magnetic flux density, temperature and 
nuclear charge $Z$, values of the ion charge in local thermodynamic equilibrium 
immediately before acceleration ($\tilde{Z}$) and at the end of the first stage of 
acceleration ($Z_{h}$) are given.  The ion charge at the end of acceleration 
$Z_{\infty}$ is given in the right-hand four columns for arbitrarily selected values of 
the general surface temperature $T_{s}$ subject to the explanation and qualifications 
discussed in Section 2.  For $T_{s} \geq 5.0\times 10^5$ K, the ion charge is always 
$Z_{\infty} = Z$.}

\begin{tabular}{@{}rrrrrrrrr@{}}
\hline
$B$ & $T_{pc}$ & $Z$ & $\tilde{Z}$ & $Z_{h}$ & $Z_{\infty}$ &
   $Z_{\infty}$ & $Z_{\infty}$ & $Z_{\infty}$    \\
$10^{12}$G & $10^{6}$K &  &  &  & $T_{s}=0.1$ & 0.2 & 0.3 & 0.4  \\
\hline
1.0 & 1.0 & 10 &  9 &  9 &      10 & 10 &    &        \\
    &     & 20 & 16 & 17 &      18 & 20 &    &        \\
	&     & 26 & 20 & 21 &      24 & 26 &    &        \\
	& 1.5 & 10 &  9 &  9 &      10 & 10 &    &        \\
	&     & 20 & 18 & 20 &      20 & 20 &    &       \\
	&     & 26 & 23 & 25 &      25 & 26 &    &        \\
	& 2.0 & 10 & 10 & 10 &      10 & 10 &    &        \\
	&     & 20 & 19 & 20 &      20 & 20 &    &        \\
	&     & 26 & 24 & 25 &      25 & 26 &    &        \\
	
3.0 & 1.0 & 10 &  7 &  7 &      8  & 10 &    &        \\
    &     & 20 & 13 & 14 &      15 & 19 & 20 &        \\
	&     & 26 & 16 & 17 &      18 & 24 & 26 &        \\
	& 1.5 & 10 &  8 &  9 &       9 & 10 &    &        \\
	&     & 20 & 16 & 17 &      17 & 19 & 20 &        \\
	&     & 26 & 19 & 20 &      20 & 24 & 26 &        \\
	& 2.0 & 10 &  9 & 10 &      10 & 10 &    &        \\
	&     & 20 & 17 & 18 &      18 & 19 & 20 &        \\
	&     & 26 & 21 & 23 &      23 & 24 & 26 &        \\
	
10.0 & 1.0 & 10 &  5 &  5 &      6 &  6 &  9 & 10     \\
     &     & 20 &  8 &  8 &     10 & 11 & 18 & 19   \\
	 &     & 26 & 11 & 11 &     12 & 13 & 22 & 24   \\
	 & 1.5 & 10 &  7 &  7 &      7 &  7 &  9 & 10     \\
	 &     & 20 & 12 & 12 &     14 & 14 & 18 & 19   \\
	 &     & 26 & 14 & 15 &     17 & 17 & 22 & 24   \\
	 & 2.0 & 10 &  8 &  8 &      9 &  9 &  9 & 10     \\
	 &     & 20 & 14 & 15 &     16 & 16 & 18 & 19   \\
	 &     & 26 & 17 & 18 &     20 & 20 & 22 & 24   \\
\hline
\end{tabular}
\end{table}

\section{Conclusions}

The results described in the previous Section are a further example of the diverse 
forms of polar-cap acceleration under the space-charge limited flow boundary condition 
that exist for the ${\bf \Omega}\cdot{\bf B} < 0$ case.
It was assumed that the composition of the accelerated plasma is uniform over the whole 
polar-cap area.  Before examining possible observable consequences of the uniform 
state, we need to mention briefly the impact of these results on the polar-cap 
acceleration which was considered at length in Paper II.

We first comment on the assumed stability of the state described in Section 2.
In the physical processes concerned, there is no obvious characteristic time apart from 
ion flight times which are of the order of $Rc^{-1} \sim 4\times 10^{-5}$ s.  This is 
many orders of magnitude shorter than the proton diffusion time constant $\tau _{p}$ 
defined in Paper II.  Thus we do not believe that the possibility of instability 
arising from photo-electric absorption can materially change the conclusions of that 
work.

Decay of the giant-dipole resonance formed in reverse-electron showers produces 
protons, and it was shown in Papers I and II that this process leads to 
temporal instabilities in the composition of the plasma accelerated. It was proposed 
that the consequences of these include the phenomena of nulls and sub-pulse drift 
observed in some radio pulsars. The number of protons created per unit ion charge 
accelerated, defined as $K$ in Papers I and II, is the significant parameter.  It was 
shown in II that, in general, the state of the polar-cap must be time-dependent, with 
any given element of area alternating between the acceleration of ions and positrons or 
of protons.  In such an element, proton acceleration produces no reverse-electron flux 
and so must be of limited duration.  It was assumed in Paper II that acceleration would 
be through 
the full potential difference given here by equations (8) and (9), but this obviously 
needs some qualification.  At any instant, there is positron and ion acceleration from 
only a fraction $(K + 1)^{-1}$ of the whole polar-cap area.  Thus the photo-electric 
production of negative charge is confined to this fraction and for $K \sim 5 - 10$ the 
charge density averaged over the cross-sectional area $\pi u^{2}_{0}(z)$ does not 
differ much from that given by equation (9).  Very roughly, the modified integrated 
form of (15),
\begin{eqnarray} 
Z_{\infty} - Z_{h} < \frac{1}{2}\kappa (K + 1)Z_{\infty},
\end{eqnarray}
is now the necessary condition for the existence of a non-zero acceleration potential 
difference.  The actual value of $Z_{\infty} - Z_{h}$ determines the amount by which 
the equilibrium acceleration potential is reduced from the full potential $\Phi$ .  
Inspection of $Z_{\infty} - Z_{h}$ values given in Table 1 shows that, for $K\geq 5$ 
and with the exception of $T_{pc} = 1.0\times 10^{6}$ K, the acceleration potential is 
usually a large fraction of $\Phi$.

The uniform state considered here in Section 2 is of interest in connection with the 
usual assumption that secondary low-energy electron-positron pair creation is necessary 
for coherent radio emission. It has no obvious mechanism for electron-positron pair 
formation and the assumption might well be that, in an isolated neutron star, it would 
correspond with a long null state of radio and polar-cap X-ray emission.

However, the possibility that cold ions might be important in the growth of unstable 
quasi-longitudinal plasma modes was considered many years ago by Cheng \& Ruderman 
(1980), but it appeared that the longitudinal effective mass $M\gamma ^{3}$ of the 
accelerated ions was too large to give the necessary growth rates. However, the small 
Lorentz factors found here prompt reconsideration because the
longitudinal effective mass is much reduced although the energy flux of the ions 
remains high in comparison with typical pulsar radio luminosities.

In the uniform composition state, acceleration fields are everywhere small so that the 
reverse-electron energy flux gives values $K < 1$ and a mixed plasma of ions and 
protons for which the instability arguments of Papers I and II are not necessarily 
valid.  The plasma then has two cold components, ions and protons with small Lorentz 
factors $\gamma _{1}$ and $\gamma _{2}$, respectively, with
$\gamma _{1} \approx 0.5\gamma _{2}$, and obviously satisfies the relativistic Penrose 
condition (see Buschauer \& Benford 1977).  But because the plasma is cold, we can 
easily find the growth rate of the quasi-longitudinal Langmuir mode studied by Asseo, 
Pelletier \& Sol (1990) for which the dispersion relation is,
\begin{eqnarray}
(\omega^{2} - k^{2}_{\parallel})\left(1 - \frac{\omega^{*2}_{1}}{(\omega - 
k_{\parallel}\beta _{1})^{2}} - \frac{\omega^{*2}_{2}}{(\omega - k_{\parallel}\beta 
_{2})^{2}}\right) = k^{2}_{\perp},
\end{eqnarray}
for angular frequency $\omega$, wave vector ${\bf k}$, and at ion and proton velocities 
$\beta _{1}$ and $\beta _{2}$, respectively. The remaining quantities are $\omega^{*} 
_{i} = \gamma^{-3/2}_{i}\omega _{i}$ in which,
\begin{eqnarray}
\omega^{2}_{i} = \frac{4\pi N_{i}q^{2}_{i}}{M_{i}},
\end{eqnarray}
where $N_{i}$, $q_{i}$ and $M_{i}$ are respectively the number density, charge and mass 
of each component $i = 1,2$.  The rest-frame plasma frequency for a given component is 
$\gamma^{-1/2}_{i}\omega _{i}$ and therefore,
following Asseo et al, we consider the case $k_{\perp} = 0$
and $k_{\parallel} = 2\omega^{*}_{1}\gamma^{2}_{1}$, defining a new variable $s$ 
through the relation,
\begin{eqnarray}
\omega - k_{\parallel}\beta _{1} = \omega^{*}_{1}(1 + s),
\end{eqnarray}
so that the dispersion relation becomes,
\begin{eqnarray}
1 - \frac{1}{(s + 1)^{2}} - \frac{C}{(s + \mu)^{2}} = 0
\end{eqnarray}
in which $\mu = \gamma^{2}_{1}/\gamma^{2}_{2}$ and $C = 
\omega^{*2}_{2}/\omega^{*2}_{1}$.  Inspection of equation (20) shows that this quartic 
has two real roots which can easily be found numerically thus giving the remaining 
complex roots.  In general, the growth rate found from the complex roots is of the 
order of $10^{-1}\omega^{*}_{1}$, but is smaller if another ion species with 
$q_{1}/M_{1} \neq q_{2}/M_{2 }$is substituted in place of the proton beam.  Thus for a 
typical ion charge of $q_{1}/e = 20$ and a proton
beam of number density $N_{2} = 5N_{1}$ we can choose $C = 0.045$ and $\mu = 0.2$, for 
which the amplitude growth rate is,
\begin{eqnarray}
{\rm Im}\omega  = \pm 0.18\omega^{*}_{1} = \pm 4.0 \times 
10^{7}\left(\frac{B_{12}R^{3}}{\gamma^{3}_{1}Pr^{3}}\right)^{1/2}
\end{eqnarray}
rad s$^{-1}$, at radius $r$, where $B_{12}$ is the polar-cap magnetic flux density in 
units of $10^{12}$ G and $P$ is the rotation period. 
The amplitude growth factor, assuming constant $\gamma _{1}$ the interval $0 < z < 
z_{max}$, is $\exp\Lambda$, where,
\begin{eqnarray}
\Lambda & = & \int^{z_{max}}_{0}dz\frac{{\rm Im}\omega}{c}  \nonumber  \\
& = & 3.2\times 10^{3}
\left(\frac{B_{12}}{\gamma^{3}_{1}P}\right)^{1/2}\left(1 - \left(\frac{R}{R + 
z_{max}}\right)^{1/2}\right).
\end{eqnarray}
This is large for $\gamma _{1} < 10$.  Values of the mode frequency are relatively 
small, $\omega \approx 2\gamma^{2}_{1}\omega^{*}_{1}$, and may lie well below $400$ 
MHz,  but it would be interesting to see if evidence for it is present in pulsar 
spectra at low frequencies.

We have seen that the presence of ion and proton beams that are relativistic but not 
ultra-relativistic, previously unexpected in pulsars, can be associated with the growth 
of unstable quasi-longitudinal modes having growth rates high enough for the formation 
of Langmuir solitons and strong turbulence.  We can refer to the review of Melrose 
(2000) and to the papers of Weatherall (1997) and (1998), Melikidze et al (2000) and 
Asseo \& Porzio (2006) to note that the collapse of such solitons is widely thought to 
be a plausible source for the coherent radio emission.  It is usually assumed that the 
instability occurs in secondary electron and positron beams with Lorentz factors of the 
order of $10^{2}$ but it is the macroscopic fields of the solitons that lead to 
coherent emission independently of the particle type from which they are formed.

\section*{Acknowledgments}

I thank the referee for directing my attention to the work of Potekhin, Pavlov \& 
Ventura (1997) on the limitations of the adiabatic approximation for the 
photo-ionization of atomic hydrogen in very strong magnetic fields.

\appendix
\section[]{Photo-electric absorption in high magnetic fields}

The presence of a high magnetic field greatly modifies the final states that are 
accessible to the electron in photo-electric absorption. In particular, the Landau 
level spacing, $\hbar\omega _{B} = 11.6$ keV at $10^{12}$ G, is large compared with 
blackbody photon energies.  In the rest-frame of the accelerated ion, the photon 
momentum ${\bf k}$ is almost parallel with ${\bf B}$ and this much reduces transition 
rates to the $n = 0$ Landau state.  We adopt the Johnson \& Lippmann (1949) spinor 
solutions for a free electron in a uniform magnetic field. These are satisfactory here 
because we require only the cross-section $\sigma _{0}$ to the unique $n = 0$ ground 
state, and the sum of the cross-sections to the two degenerate $n = 1$ states.  Photon 
polarization states are defined as those perpendicular to, or parallel with, ${\bf 
k}\times {\bf B}$ and are denoted by superscripts $(a,b)$ respectively.  For the 
explicit form of the solutions and of the interaction with the electromagnetic field  
it will be convenient here to refer to Appendix A of Jones (2010b).

Although with later comment, we assume the adiabatic approximation for the 
partially-ionized ground state in which the electrons occupy single-particle states 
each with Landau quantum number $n = 0$.  Recoil is neglected so that in
the ion rest frame, the photo-electric thresholds are $\epsilon _{l} + n\hbar\omega 
_{B}$, where $\epsilon _{l}$ is the separation energy given by equation (1) and $n$ is 
here the electron final state Landau quantum number.
Following equation (A6) of Jones (2010b), the
transition matrix element from a photon with angular frequency $\omega$ and 
polarization $\mbox{\boldmath$\epsilon$}^{a,b}$ to the
$n = 0$ ground state with electron longitudinal momentum $p$ is,
\begin{eqnarray}
\left(\frac{2\pi \hbar e^{2}c^{2}}{\omega}\right)^{1/2}\int r_{\perp}dr_{\perp}d\phi
{\rm e}^{{\rm i}{\bf k}_{\perp}\cdot{\bf r}_{\perp}}  \nonumber  \\
\Psi^{\dagger}_{-1,0}(p)\mbox{\boldmath$\alpha$}\cdot\mbox{\boldmath$\epsilon$}
^{a,b}
\Psi _{-1,0}(p - k_{\parallel})G(p - k_{\parallel})
\end{eqnarray}
in cylindrical polar coordinates,  where $G$ is the Fourier transform of the model 
bound electron state $\sqrt{\kappa}\exp(-\kappa|z|)$, in which $\hbar^{2}\kappa^{2} = 
2m\epsilon _{l}$.  The dependence of the cross-section on the charge of the ion is 
present solely through the Fourier transform. The model function, which replicates the 
correct asymptotic form of the true function, given in this Appendix by the parameter 
$\kappa$, is an adequate approximation for the purposes of this paper. 
The Johnson \& Lippmann spinor for Landau number $n \geq 0$ is,
\begin{eqnarray*}
\Psi _{-1.n}(p) = \left[\begin{array}{c}  0    \\
\psi _{n}   \\
\frac{p_{\perp}}{2mc}\psi _{n-1}    \\
-\frac{p}{2mc}\psi _{n}
\end{array}\right],
\end{eqnarray*}
where $p^{2}_{\perp} = 2mn\hbar\omega _{B}$, and the normalized functions
$\psi _{nl}$ are the non-relativistic radial solutions of the Schrodinger equation for 
a free electron in a uniform magnetic field, in which $l$ is the spatial degeneracy 
quantum number (see equations A3 - A5 of Jones 2010b).  The azimuthal angle-dependence 
is $\exp(-{\rm i}(l - n)\phi)$.  The ion structure consists of spatially well-defined 
coaxial shells with sequential values $l = 0, 1, 2....$ and energies given by equation 
(1).  The possibility of a hole state is not considered because its lifetime would be 
many orders of magnitude smaller than the photo-electric transition rates calculated 
here.

For the specific geometry of the photo-electric  process considered here and for high 
fields $B \geq 10^{12}$ G, the exponent in the matrix element, which is of the order of 
$k_{\perp}r_{B}$ is always very small compared with unity and is set equal to zero. 
This introduces the selection rule $\delta(n - l) = 0$ which  leads to an easy 
evaluation of the transition rates. Hence, immediately from equation (A1), we obtain 
the cross-section,
\begin{eqnarray}
\sigma^{a}_{0} = \frac{2\pi e^{2}\sin^{2}\chi}{cpm\omega} \hspace{4cm} \nonumber  \\
\left(\frac{\kappa^{3}(2p - k_{\parallel})^{2}}
{(\kappa^{2} + (p - k_{\parallel})^{2})^{2}} +    
\frac{\kappa^{3}(2p + k_{\parallel})^{2}}{(\kappa^{2} + 
(p + k_{\parallel})^{2})^{2}}\right),
\end{eqnarray}
for transitions to the $n = 0$ Landau state for polarization perpendicular to ${\bf 
k}\times {\bf B}$.  In the parallel polarization case, the cross-section to the $n = 0$ 
state is $\sigma^{b}_{0} = 0$.  Proceeding in a similar way, the cross-sections to the 
two degenerate $n = 1$ states are found to be,
\begin{eqnarray}
\sigma^{a}_{1} = \frac{2\pi e^{2}\cos^{2}\chi}{cpm\omega} \hspace{4cm}
\nonumber  \\
\left(\frac{\kappa^{3}(p^{2}_{\perp} + k^{2}_{\parallel})}{(\kappa^{2}
 + (p - k_{\parallel})^{2})^{2}} + 
 \frac{\kappa^{3}(p^{2}_{\perp} + k^{2}_{\parallel})}
 {(\kappa^{2} + (p + k_{\parallel})^{2})^{2}}\right),
 \end{eqnarray}
and $\sigma^{b}_{1} = \sigma^{a}_{1}$.  It is to be emphasized again that these 
expressions are valid only under the condition that $k_{\perp}r_{B} \ll 1$ which is 
valid for the photo-electric process considered here. These cross-sections, not too far 
above threshold and at $10^{12}$ G, do not differ greatly from those given by the 
zero-field Kramers formula apart from the presence of the $\sin^{2}\chi$ factor in 
$\sigma^{a}_{0}$, and the zero parallel-polarization cross-section $\sigma^{b}_{0}$.  
These factors, within the adiabatic approximation, are a direct consequence of the 
relative orientation of the photon polarization vector and the magnetic field.  
Transitions from $n = 0$ to final states of $n \geq 2$ have not been considered here 
because their matrix elements vanish in our approximation $k_{\perp}r_{\perp} = 0$.

The procedures used here are much inferior to those for zero-field cross-sections as in 
the work of Reilman \& Manson (1979), in particular, the  unperturbed final-state wave 
function and the state of the initial bound electron.  The latter has been treated in 
the adiabatic approximation as a single-particle state with good quantum numbers $n = 
0$ and $l$.  Use of this approximation in the case of atomic hydrogen has been 
extensively studied by Potekhin, Pavlov \& Ventura (1997).  The true ground state 
contains small components with $n > 0$ which lead to a non-zero value of the 
cross-section $\sigma^{b}_{0}$ which increases in significance as the threshold for the 
$n = 1$ final state is approached.  The same is true for any single-particle state in a 
multi-electron ion.  But in multi-electron ions, there are further problems. Whilst a 
shell model is clearly the only useful basis for description of the atom, as in the 
zero-field case, there must be a residual electron-electron interaction whose matrix 
elements, though satisfying the $\delta(n - l) = 0$ selection rule overall, can 
introduce an $n = 1$ and $l + 1$ component into the $n = 0$ and $l$ ground state and 
partially invalidate the adiabatic assumption.
Relative to unity, the amplitude of such a component would be of order 
$\Delta/\hbar\omega _{B}$, where $\Delta$ is the matrix element, and is almost 
certainly very small. But the component has a photo-electric transition matrix element 
to the $n = 0$ final electron state that is not inhibited by the $\sin\chi$ factor.  
Thus there must be a small non-zero value of
$\sigma^{b}_{0}$ and a correction to $\sigma^{a}_{0}$.  Clearly the cross-sections 
obtained here are subject to some uncertainty for the many reasons that have been 
discussed. However, we do not need great accuracy in the present work because the 
photo-electric transition rates required are also dependent on the high-frequency 
exponential tail of the blackbody spectrum and on the Lorentz transformation to the ion 
rest frame.  But it is important that  cross-sections with the correct high-field 
Landau level structure and dependence on photon polarization be used.

\bsp

\label{lastpage}


\begin{thebibliography}{99}
\bibitem{b1}Asseo E., Pelletier G., Sol H., 1990, MNRAS, 247, 529
\bibitem{b2}Asseo E., Porzio A., 2006, MNRAS, 369, 1469
\bibitem{b3}Buschauer R., Benford B., 1977, MNRAS, 179, 99
\bibitem{b4}Cheng A. F., Ruderman M. A., 1977, ApJ, 214, 598
\bibitem{b5}Cheng A. F., Ruderman M. A., 1980, ApJ, 235, 576
\bibitem{b6}Goldreich P., Julian W. H., 1969, ApJ, 157, 869
\bibitem{b7}Harding A. K., Muslimov A. G., 2001, ApJ, 556, 987
\bibitem{b8}Johnson M. H., Lippmann B. A., 1949, Phys. Rev., 76, 828
\bibitem{b9}Jones P. B., 1981, MNRAS, 197, 1103
\bibitem{b10}Jones P. B., 1985, MNRAS, 216, 503
\bibitem{b11}Jones P. B., 2010a, MNRAS, 401, 513
\bibitem{b12}Jones P. B., 2010b, MNRAS, 409, 1719
\bibitem{b13}Jones P. B., 2011, MNRAS, 414, 759
\bibitem{b14}Medin Z., Lai D., 2006, Phys. Rev. A., 74, 062507
\bibitem{b15}Melikidze G. I., Gil J. A., Pataraya A, D., 2000, ApJ, 544, 1081
\bibitem{b16}Melrose D. B., 2000, ASP, 202, 721
\bibitem{b17}Michel F. C., 1974, ApJ, 192, 713
\bibitem{b18}Mori K., Ho W. C. G., 2007, MNRAS, 377, 905
\bibitem{b19}Muslimov A. G., Tsygan A. I., 1987, Astrophysics, 29, 625
\bibitem{b20}Muslimov A. G., Tsygan A. I., 1992, MNRAS, 255, 61
\bibitem{b21}Muslimov A. G., Harding A. K., 1997, ApJ, 485, 735
\bibitem{b22}Potekhin A. Y., Pavlov G. G., 1997, ApJ, 483, 414
\bibitem{b23}Potekhin A. Y., Pavlov G. G., Ventura J., 1997, A\&A, 317, 618
\bibitem{b24}Reilman R. F., Manson, S. T., 1979, ApJ Suppl., 40, 815
\bibitem{b25}Ruderman M. A., Sutherland P. G., 1975, ApJ, 196, 51
\bibitem{b26}Wang N., Manchester R. N., Johnston S., 2007, MNRAS, 377, 1383
\bibitem{b27}Weatherall J. C., 1997, ApJ, 483, 402
\bibitem{b28}Weatherall J. C., 1998, ApJ, 506, 341
\bibitem{b29}Yakovlev D. G., Pethick C. J., 2004, Ann. Rev. Astron. Ap., 42, 169

\end{thebibliography}
\end{document}